\documentclass[journal]{src/vgtc}                     

\onlineid{6658}

\vgtccategory{Research}

\vgtcpapertype{Application}

\author{%
  \authororcid{Wong Kam-Kwai}{0000-0002-2813-1972},
  \authororcid{Yi-Lin Ye}{0009-0000-9986-4420},
  \authororcid{Wai Tong}{0000-0001-9235-6095},
  \authororcid{Haobo Li}{0009-0003-8771-2835},
  \authororcid{Kentaro Takahira}{0009-0003-5613-610X},
  \texorpdfstring{\\}{}
  Aastha Bhatta,
  Sunil Poudyal,
  Charles Wang Wai Ng,
  \authororcid{Huamin Qu}{0000-0002-3344-9694}, and
  \authororcid{Leni Yang}{0000-0003-4527-4905}
}

\authorfooter{
  \item
  	W. Kam-Kwai, Y. Ye, H. Li, K. Takahira, and H. Qu are with Dept of Computer Science and Engineering of HKUST. 
    Email: \{kkwongar, yyeaz, hliem, ktakahira\}@connect.ust.hk and huamin@cse.ust.hk
  \item W. Tong is with Texas A\&M. Email: wtong@tamu.edu.
  \item
  	A. Bhatta, S. Poudyal, and CWW Ng are with Dept of Civil and Environmental Engineering of HKUST.
  	Email: \{abhatta, spoudyal\}@connect.ust.hk and charles.ng@ust.hk.
  \item
    L. Yang (corresponding) is with Inria, CNRS. Email: leni.yang@inria.fr.
}

\title{LandSAR: Visceralizing Landslide Data for Enhanced\\ Situational Awareness in Immersive Analytics}

\abstract{
Landslides pose a significant threat to public safety, but their dynamic processes are difficult to analyze from post-event observation alone. 
Computational simulation is therefore essential, but it generates vast, abstract datasets that create a cognitive gap between the analyst and the real-world, physical terrain. 
While Immersive Analytics (IA) begins to bridge this gap by visualizing data in 3D, we explore how these systems evolve beyond abstract data and integrate data visceralization to enhance Situational Awareness (SA).
We present LandSAR, an immersive analytics system that enhances SA for landslide analysis by visceralizing landslide data through integrated simulations and visualizations.
LandSAR supports real-time simulations of landslide dynamics, prevention strategies, and climate impacts, enabling multi-perspective what-if analyses.
The system uses 3D-printed terrain models as tangible interfaces to facilitate haptic feedback and enable gesture-based exploration, allowing for intuitive geographical perception.
Expert interviews and workshops demonstrate that LandSAR effectively improves SA and engagement.
}


\teaser{
  \centering
  \includegraphics[width=\linewidth, alt={The image illustrates LandSAR, an augmented reality (AR) system designed to enhance landslide awareness using a physical terrain model. In Panel A, AR overlays highlight landslide risks directly on the model. Panel B shows a person wearing an AR headset, interacting with the model through touch, making it intuitive to explore potential hazards. In Panel C, a first-person view of a virtual landscape simulates a hypothetical landslide, immersing the user in a realistic experience that helps visualize and understand the risks involved. This combination of AR and physical interaction offers a comprehensive approach to landslide risk assessment.}]{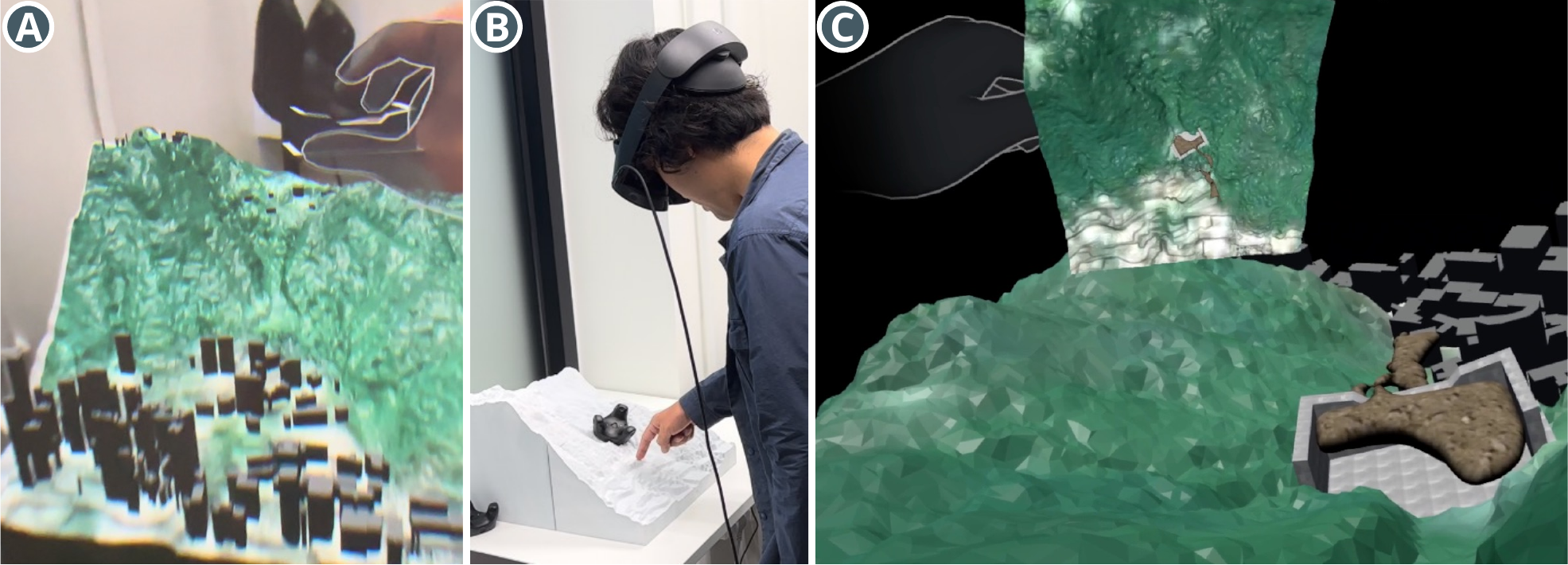}
  \caption{%
  	LandSAR situates visualizations of landslides on tangible terrain models. 
    It enhances situational awareness during landslide events through three main features: (A) augmented reality for situated visualizations highlighting landslide risks, (B) physical interactions on terrain models for intuitive use, and (C) a first-person perspective for an immersive, hypothetical landslide experience.%
  }
  \label{fig:teaser}
}

\graphicspath{{figs/}{figures/}{pictures/}{images/}{./}} 

\usepackage{tabu}                      
\usepackage{booktabs}                  
\usepackage{lipsum}                    
\usepackage{mwe}                       

\usepackage[svgnames, dvipsnames]{xcolor}

\usepackage{mathptmx}                  

\newcommand{\rc}[1]{#1}
\newcommand{\rw}[1]{#1}

\usepackage{xspace,xpunctuate}

\newcommand{\ie}{\textit{i.e.},\xspace}
\newcommand{\etal}{\xspace\textit{et al.}\xspace}
\newcommand{\eg}{\textit{e.g.},\xspace}

\begin{document}


\maketitle
\section{Introduction}
Landslides are serious hazards that inflict substantial fatalities and economic losses globally every year.
These threats are expected to intensify as climate change increases the frequency and intensity of rainfall, which is one of the primary triggers for landslides~\cite{intro_motivation_extreme_2019}.
To develop effective mitigation strategies and predict risks, understanding these complex spatio-temporal events remains a major challenge in geoscience and civil engineering. 
Landslide behavior is highly context-dependent, and its critical processes of failure, such as rapid soil saturation, occur chaotically and are often only observable in their aftermath~\cite{intro_motivation_landslide_2014}.
Traditional post-hoc analysis, which relies on empirical and analytical models, provides only a static representation of the event's outcome but not the dynamic processes~\cite{intro_challenge_assessment_2002}. 
Consequently, modern analysis has shifted toward computational simulation to approximate these dynamics which are difficult to observe.

However, this simulation-based approach introduces a sense-making challenge caused by data abstraction. 
Simulations generate vast, high-dimensional, and time-varying datasets. 
When explored through conventional 2D charts like heatmaps and contour plots, the data remains abstract and symbolic. 
Analysts have to bear a substantial cognitive load to mentally reconstruct a complex 4D (3D space + time) physical process (\ie landslide mechanisms) from these abstractions.
This cognitive gap impedes intuition and hinders the achievement of Situational Awareness (SA). 
In this context, we adopt Endsley's definition of SA~\cite{related_sa_endsley_1995} as the analyst's capacity to (1) perceive the current state of the terrain, (2) comprehend the mechanism of its dynamic processes, and (3) project its status into the near future.
The cognitive gap described above represents a failure at Level 2 (Comprehension) and Level 3 (Projection) SA, as analysts struggle to mentally connect the static, symbolic data with dynamic physical processes.

Immersive analytics~\cite{immersive_analytics} has emerged as a promising solution to bridge this cognitive gap, using 3D representations to support the analysis of high-dimensional, spatio-temporal data by providing extended virtual space and improved spatial reasoning~\cite{related_immersive_filho_2020, related_immersive_miria_2021}.
However, its potential is not limited to transitioning analysis from 2D screens to extended virtual spaces. 
According to theories of embodied cognition, human understanding of physical phenomena is rooted in sensory-motor experiences~\cite{embodied_congition_2008}. 
For example, we understand concepts like ``flow,'' ``slope,'' and ``friction'' through a kinaesthetic intuition developed over a lifetime of interacting with the physical world. 
Yet, most immersive analytics systems focus on 3D representations of abstract data, without helping analysts connect that data to their kinaesthetic intuition.
Data remain symbolic, disembodied, and disconnected from the user's kinaesthetic understanding, which is insufficient for enhancing spatial sense-making~\cite{intro_motivation_ens_2022} and situational awareness~\cite{related_visceral_diagram_2025} required for complex tasks like landslide analysis.

To bridge this disembodied gap, we explore the synthesis of two key concepts: \textit{data physicalization}~\cite{related_tangible_physicalization_2015} and \textit{data visceralization}~\cite{related_visceralization_2020}. 
First, data physicalization uses physical artifacts as the medium for data representation and interaction. 
This integrates haptic feedback and grounds the user's perception~\cite{related_tangible_herman_2021}. 
Second, simulations of physical processes can act as a form of data visceralization. Data visceralization restores the intuitive understanding of physical measures by allowing users to experience dynamic behaviors rather than viewing static abstractions~\cite{related_visceralization_2020}.
By showing the landslide's dynamic behaviors (\eg flow velocity and impact force) rather than static visualizations, the simulation provides an experiential understanding of the data. 
This allows analysts to apply their kinaesthetic intuition to gain a qualitative understanding of the physical measures. 
However, this introduces a trade-off between two analytical modes.
On the one hand, the analytical mode of immersive visualization relies on the cognitive processing of symbolic overlays. 
On the other, the intuitive mode of simulation-based visceralization relies on an embodied understanding of physical processes.
This leads to our core research question: how can the rational, analytical mode and the intuitive, embodied mode be effectively synthesized to create a holistic analytical environment that enhances all three levels of analyst situational awareness for landslide analysis?

We present LandSAR, an immersive analytics system that enhances SA for landslide analysis by visceralizing landslide data through integrated simulations and visualizations.
To effectively bridge the analytical and intuitive modes, LandSAR contains three core modules:
\textbf{Immersive visualization as the analytical base:} LandSAR uses AR-based situated visualizations to project historical data, geographical context, and risk predictions onto physical terrain proxies~\cite{related_tangible_proxsituated_2023}, leveraging the user's analytical knowledge and providing the cognitive entry point to the data.
\textbf{Simulation-based visceralization for dynamic feedback:} The system integrates a computational steering pipeline to visceralize physical phenomena. 
Users can conduct what-if analyses of mitigation strategy designs and interpret the dynamic consequences immediately. 
This restores a qualitative understanding of the physical measures (Level 1 SA), allowing them to test if the landslide's behavior matches their kinaesthetic intuition (Level 2 SA) and thus improve their predictive comprehension (Level 3 SA).
\textbf{Data physicalization as the embodied medium:} We use 3D-printed terrain models as a form of data physicalization and the primary medium for interaction. 
This tangible interface provides haptic feedback to reduce cognitive load and enable accurate spatial judgments~\cite{intro_motivation_designar_2019}, acting as a stable physical anchor for the virtual overlays. 
Through an iterative co-design process and subsequent expert evaluation workshops, we assessed how this synthesis of components impacts the analytical process and enhances SA.
In summary, our main contributions are:
\begin{itemize}[noitemsep,topsep=0pt,label=$\diamond$]

\item A problem characterization of landslide risk analysis mapped to the three-level Situational Awareness framework. We formulate simulation-based visceralization as a component for enhancing the comprehension (Level 2) and projection (Level 3) levels of SA in immersive analytics.

\item The design and implementation of LandSAR, an immersive analytics system that synthesizes immersive visualization, simulation-based visceralization, and data physicalization as an evaluation probe for the disaster management domain.

\item Empirical findings and design implications from our expert evaluation, providing insights into how the synthesis of analytical and intuitive modes can be effectively designed to enhance all three SA levels in spatio-temporal analytical tasks.

\end{itemize}
\section{Related work}
Our work focuses on immersive data visualization in urban settings, relating to literature that similarly aims to enhance SA, addresses spatial-temporal data, and employs tangible interfaces for interaction.

\subsection{Immersive visceralization for situational awareness}
Situational awareness (SA) is commonly conceptualized in three hierarchical phases: perceiving relevant elements, comprehending them, and projecting their future states~\cite{related_sa_endsley_1995}.
Many visual analytic systems aim to enhance SA in time-sensitive applications and high-stakes decision-making (\eg emergency management~\cite{related_sa_bosch_2013, related_sa_snyder_2020, haobo2025havior}).
To capture the temporal dynamics of these environments, such systems often prioritize high-velocity data streams, relying heavily on textual sources (\eg social media and news) to ensure timeliness. 
Yet, they frequently lack deep spatial integration.
In spatial contexts like urban planning, visualization interfaces have evolved to address geographical complexity through situated visualizations~\cite{related_sa_moere_2012, related_sa_public_survey_2021, yang2025augmented}, multi-display systems~\cite{related_tangible_cityscope_2018, related_tangible_uplift_2021}, and cross-virtuality platforms~\cite{related_sa_public_2022}. 
These approaches typically focus on delivering static information for perceiving and comprehending long-term urban issues, but they often lack adequate support for dynamic situations. Furthermore, their reliance on abstract representations can disconnect the visuals and the underlying meaning of the data.

The concept of \textit{data visceralization} has been proposed by Lee\etal~\cite{related_visceralization_2020} to address these limitations. 
Beyond seeing data in abstract encodings, it seeks to represent the data intuitively so that users can feel its meaning physically and emotionally.
Although this approach can improve sense-making~\cite{related_visceral_diagram_2025}, less attention has been paid to its application to complex and procedural environmental hazards like landslides. AR provides an immersive medium that allows the creation of dynamic first-person or third-person experiences in real time and transforms an abstract risk into an impactful event and enhances SA~\cite{related_sa_whitlock_2020, related_sa_woodward_2023, related_visceral_experience_2024}. 
For example, presenting an annotated World-In-Miniature of the environment~\cite{related_sa_ar_2002} offers a broader spatial context while replaying asynchronous causal events~\cite{related_sa_causality_2022} mitigates temporal inconsistencies due to missing information.
Inspired by this potential, our work extends the concept of visceralization by using AR-based real-time simulation to visceralize the entire landslide process, transforming an abstract risk into a dynamic and impactful event that users can experience directly. This approach bridges the spatial and temporal gaps that impede users' SA in dynamic landslide scenarios, enabling them to assess risks in landslide-prone areas more effectively.


\subsection{Immersive analytics for spatial-temporal data}
Immersive Analytics (IA) explores novel methods for visualizing and interpreting data through immersive experiences, as highlighted by recent surveys~\cite{related_immersive_survey_CVA_2022, related_immersive_survey_grand_2021, related_immersive_survey_3d_2022, related_immersive_survey_IA_2021}.
With a greater integration with reality, the immersive experience can augment human capabilities to levels reminiscent of fictional superpowers~\cite{related_immersive_pie_2022}.
Our work empowers users to visualize rare emergency scenarios (\ie landslide) and enhance their situational awareness in such events.

IA provides new perspectives to overcome challenges in visualizing spatial data, which is inherently sampled from 3D environments but is typically presented in 2D due to 3D occlusion issues~\cite{related_immersive_tory_2006}.
Cartograms, which depict real-world geographic contexts, have attracted considerable interest in enriching their design space, particularly in terms of interactive techniques.
Ghaemi\etal~\cite{related_immersive_proxemic_2022} explored proxemic interactions to alter the underlying geometry of immersive maps.
Newbury\etal~\cite{related_immersive_newbury_2021} designed embodied gesture interactions (\eg push and pull) to perform visualization tasks on origin-destination flow data.
Yang\etal~\cite{related_immersive_tilt_2021} utilized controller tilting angles to morph area-linked data into 2D choropleths, 3D prism maps, and bar charts.
Beyond serving as standalone components, cartograms can offer spatial context to other visualizations when placed adjacently~\cite{related_immersive_filho_2020}.
These studies demonstrate that interactions are crucial for addressing occlusion problems and enhancing spatial understanding in immersive environments.

Providing spatial contexts enables in-situ analysis and the recreation of original settings to bridge spatial gaps.
This approach has proven valuable for examining trajectory data in real-world~\cite{related_immersive_timetables_2022}, abstracted physical environments~\cite{lima2025dataseum, tao2025egoexo}, and virtual environments~\cite{related_immersive_miria_2021}.
Notable examples include TIVEE~\cite{related_immersive_tivee_2022}, which created small multiples of badminton courts for analyzing tactics in badminton trajectories, and ShuttleSpace~\cite{related_immersive_shuttlespace_2021}, which displayed 2D and 3D information about badminton trajectories in an immersive court setting.
These systems vary in data representation designs on similar datasets, offering first- and third-person perspectives.
Such diverse design requirements showcase the vast design space for immersively analyzing spatial-temporal data.
Unlike many approaches that solely employ Virtual Reality (VR) as the immersive device, we explore the combination of AR and VR to harness both benefits~\cite{tong2023asymcollab}.
We support AR interactions with tangible terrain models for landslide simulation to provide intuitive realism, enable tangible interactions, and enhance embodied sensemaking~\cite{intro_motivation_ens_2022}.


\subsection{Tangible proxies for situated visualization}
Immersive visualizations offer new ways to understand data, but it is still challenging to provide a seamless experience in transforming between 2D and 3D environments~\cite{related_tangible_lee_2022}. 
On the other hand, situated visualization connects visualizations with physical environments and referents, which has been widely used in engineering and civic applications to serve users who require strong spatial understanding and engagement with the places~\cite{related_tangible_bressa_2022}.
Direct coordination between the physical world and visualization reduces the cognitive load when users explore multidimensional data~\cite{related_tangible_cordeil_2017}.
Tangible interfaces, built upon physical interactions, have emerged as a viable approach to conducting interactive queries and exploration in immersive environments~\cite{related_tangible_dane_2012}.

The visualizations used for tangible interfaces are mainly situated and embedded visualization~\cite{related_tangible_embedded_2017}.
The Embodied Axes~\cite{related_tangible_cordeil_2020}, explicitly designed for data visualizations, consist of three orthogonal axes that function as real-world sliders.
Tong\etal~\cite{related_tangible_paper_2023} extensively studied the interaction design space of printed data visualizations.
These interfaces, however, are created without real-world meaning and are entirely detached from the application, similar to a mouse and keyboard~\cite{tong2025hybrid, kentaro2025tangiblenet}.
While these tools enable physical interactions, they still lack physical contexts. 
Fleck\etal~\cite{related_tangible_ragrug_2022} noted that the main differences between immersive and situated analytics lie in the absence of physical context, interaction, and embedded perception of virtual content.

To incorporate spatial meaning into interfaces, Satriadi\etal~\cite{related_tangible_satriadi_2022} examined scale models and explored view arrangement and chart layout for visualizing multivariate data and multiple charts.
In subsequent works, they utilized tangible globes~\cite{related_tangible_globe_2022} to demonstrate the power of physical models.
Further, they proposed the ProxSituated visualization~\cite{related_tangible_proxsituated_2023}: a physical proxy for environments can bridge the spatial and temporal discrepancies.
This property is particularly useful in urban settings~\cite{related_tangible_hull_2017, related_tangible_cityscope_2018}, frequently using real-world scale models~\cite{related_tangible_miniature_2021}.
For example, Roo and Hachet~\cite{related_tangible_reality_2017} created volcano mock-ups with sandboxes and VR to dissect complex relationships.
Terrain mockups projected with geographical information were used to teach terrain analysis~\cite{related_tangible_millar_2018}.

Taking inspiration from these works, our research focuses on utilizing tangible terrain models instead of placing buildings on top of physical displays.
Compared to tabletop systems, terrain models may not support different levels of detail and changing locations as effectively, but they offer higher overall mobility, tangible depth perception, and haptic sense support, potentially enhancing interactive data exploration tasks~\cite{related_tangible_herman_2021, related_tangible_herman_2025}. 
Moreover, buildings are more likely to change than the terrain surfaces~\cite{related_tangible_bladin_2018}, so representing them virtually enhances sustainability of data physicalization~\cite{discussion_physicalization_sustainability_2024}.
In this paper, we explore using tangible 3D environmental proxies with real-world spatial meaning to interact with situated visualizations and provide an embedded perception of landslide risks.
This approach offers users a seamless and intuitive experience that bridges the gap between the physical and virtual worlds, ultimately enhancing their understanding of complex landslide scenarios.

\section{Background and design requirements}

This section describes the design process of LandSAR, which involves several discussions with domain experts and iterative development cycles.
The domain abstraction provides essential context to understand the complexity of our target domain.

\subsection{Landslide}
\begin{figure}[t]
    \centering
    \includegraphics[width=\linewidth, alt={The figure illustrates a mountain where a landslide has occurred, highlighting five key components: rainfall, landslide, slope, risk, and rigid barrier. Rainfall intensity and duration influence the event. The slope’s angle, soil type, and vegetation affect landslide occurrence. The landslide progresses through initiation, propagation, and runout stages. Risks are primarily to infrastructure, buildings, and people. Rigid barriers are used to mitigate risks in high-risk areas by controlling the landslide’s impact and overflow, requiring engineers to consider impact mechanisms and overflow dynamics for effective protection.}]{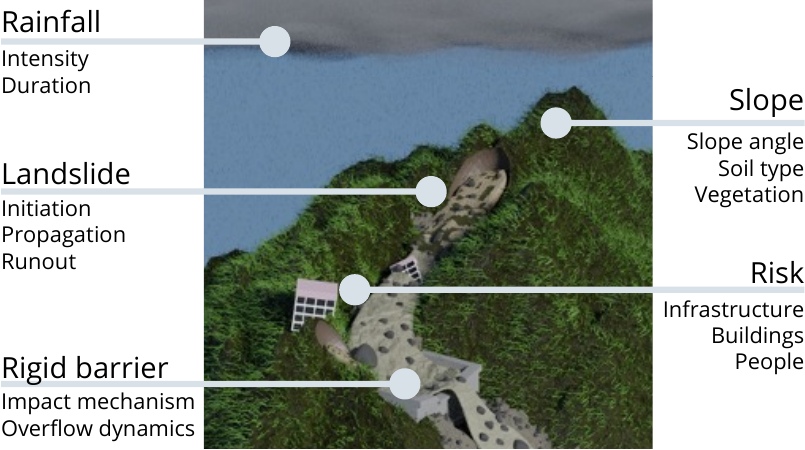}
    \caption{Illustration for the main components in landslide risk. Rainfall, which is influenced by intensity and duration; Slope, defined by its angle, soil type, and vegetation, impacting landslide risk; Landslide, which goes through initiation, propagation, and runout stages; Risk, primarily affecting infrastructure, buildings, and people in the landslide's path; and Rigid barrier, a mitigation structure that engineers use to manage the impact and overflow dynamics in high-risk areas.}
    \label{fig:landslide}
\end{figure}

Landslides are geological phenomena referring to the downward rock and soil movements due to gravity.
They occur under diverse environmental conditions~\cite{background_environmental_evis_2022}, which result in their respective forms~\cite{background_landslide_2014}.
This study specifically focuses on debris flows, a fast-moving flow-like landslide containing a mixture of \textit{boulders} (large rocks) and \textit{debris} (water, mud, aggregates, and other substances).
The key components of landslides are outlined in \cref{fig:landslide}:
\begin{itemize}[nolistsep]
\item \textit{Rainfall} is one of the primary triggers of landslides, particularly for debris flows.
When rainwater seeps into the soil, it reduces the stability of the slope.
Consequently, rainfall intensity and duration are key factors in predicting a slope's susceptibility to landslides~\cite{background_landslide_rainfall_2021}.
Weather forecasts have facilitated landslide early warning systems based on the predicted rainfall intensity and duration in vulnerable areas.

\item \textit{Slopes} vary in their physical and geological properties (\eg slope angle and soil type), determining their landslide susceptibilities. 
Steeper slopes are more susceptible due to the stronger driving forces of gravity, which overcome the frictional strength of the slope material. 
Vegetation can provide additional slope stability by reinforcing the soil with root systems. 

\item \textit{Landslides} generally progress through three stages: initiation (where and how it starts), propagation (how it flows), and runout (how far it reaches). 
Investigating landslide initiation helps geotechnical engineers design slope protection measures (\eg vegetation~\cite{intro_motivation_landslide_2014}) to prevent landslides from occurring. Meanwhile, studying propagation and runout helps with landslide risk assessments, which inform decision-making in finding economical solutions to reduce the potential impacts on life and property.

\item \textit{Rigid barriers} are preventive measures that mitigate landslide \textit{risks} by reducing its propagation and runout. 
However, barriers can be damaged or overflown if the landslide's impact or volume exceeds design capacity, leading to further propagation~\cite{visual_velocity_2021}.
\end{itemize}

\subsection{Co-design with domain experts}
\label{sec:interview}
Inspired by the study methodology of Uplift~\cite{related_tangible_uplift_2021}, we adopted the Co-design procedure~\cite{design-probes} to tackle the complex domain in landslide risk education.
Three co-authors from non-computer science backgrounds joined the development of LandSAR, contributing domain-specific knowledge and helping evaluate the system from an interdisciplinary perspective.
The co-design process was initiated as part of a research project to develop immersive educational tools for landslide awareness. 
While the initial stages focused on public education, these discussions with domain experts revealed a deeper, more pressing need in expert-level sense-making. 
The primary goal shifted from scientific communication for the public to facilitating expert comprehension and projection (Levels 2 and 3 SA).

To investigate this gap, we conducted weekly meetings over six months to iteratively design and deploy a technology probe~\cite{background_technology_probe_2003}.
This probe was evaluated by domain experts ($N=9$) and senior geotechnical engineers from the local government ($N=2$) with over 15 years of professional experience.
This formative study allowed us to understand the limitations of conventional immersive approaches and identify the core requirements for a system capable of supporting expert-level SA.

\subsection{Formative findings}
The technology probe provided several key insights that pivoted the project's focus from public education toward expert analysis, guiding the subsequent design of LandSAR.

\textbf{Finding 1: mid-air interaction creates a disembodied gap.} 
The probe allowed users to interact with virtual objects and immersive visualizations through freehand gestures. 
Previous studies, such as Satriadi\etal~\cite{object_interaction_satriadi_2019}, showed that freehand gestures could reduce fatigue and perform adequately for tasks like map transitions.
However, our pilot tests revealed a critical issue: difficulty in perceiving the depth of virtual objects. 
Many users struggled to reach for objects in front of them, making repetitive uncertain movements. 
Also, some remained stationary, waiting for a visual or haptic ``trigger'' to prompt them for interaction.
This reliance on purely visual, mid-air interaction created a disembodied experience, which \rc{caused spatial ambiguity and imprecise input that} significantly impeded their ability to navigate. 
This finding motivated the need for a physical, tangible anchor to ground perception and interaction. It motivates our data physicalization module.
In addition, we redesigned the system to separate map navigation from object interaction, allowing users to focus more clearly on individual elements without becoming disoriented.

\rc{
\textbf{Finding 2: experts require parameter manipulation to build trust.} 
We initially used pre-rendered landslide simulation playbacks to illustrate landslide processes. 
However, we found that this passive approach has limited value because it prevented experts from validating the simulation against their professional intuition.
For example, several users expressed interest in seeing simulations of landslides occurring in their local areas, with the ability to adjust factors such as the size and material of rigid barriers or rainfall intensity. 
This helps them test specific hypotheses and verify if the simulated consequences matched their geotechnical expectations.
Without these options, the experience was reduced to passive observation, preventing users from gaining a deeper and more active understanding of landslide mechanics. 
As a result, the use of AR HMDs felt underwhelming and unmotivated, with the immersive potential largely unused.
This issue became more evident when domain experts had to rely on verbal explanations to convey hypothetical scenarios. 
Passive observation was insufficient for building deep comprehension (Level 2 SA). 
This finding indicates that for expert analysis, interactivity is not merely an engagement feature but also a functional requirement for computational steering. Users should be able to conduct real-time ``what-if'' analyses to bridge the gap between their internal mental models and the external physical process.}

\textbf{Finding 3: experts require more than simulations.} 
The probe offered generic and pre-scripted advice for the public. 
During evaluations, domain experts expressed a clear need for a tool that would allow them to dynamically query real-world events and validate hypothetical scenarios based on the simulation. 
They needed to see the analytical data (\eg rainfall intensity, soil type) for susceptibility analyses and simultaneously observe its consequences. 
This highlighted the core research challenge: experts require more than just a data-query tool or a simulation tool; they need a synthesis of the two. 
They need to seamlessly move between the analytical mode (symbolic data) and the intuitive mode (concrete simulation) to build a holistic understanding.

\subsection{Design requirements}
In view of these formative findings, we summarized and discussed the design requirements iteratively to derive four design considerations for expert analysis.

\begin{enumerate}[label=\textbf{R{\arabic*}}]
\item \textbf{Bridge the disembodied gap with a tangible interface.} To combat the cognitive load and perceptual ambiguity of mid-air interactions (Finding 1), the system should be grounded on a physical, tangible artifact. This data physicalization should act as a stable haptic anchor for AR overlays, supporting Level 1 SA (Perception).
This could involve using intuitive physical interactions, clear visual cues, and eliminating the need for extensive instructions or training for users without prior experience interacting with spatial displays.

\item \textbf{Enable intuitive comprehension via simulation-based visceralization.} To move beyond the failure of passive playback (Finding 2), the system should provide an experiential understanding of the simulation dataset.
It should incorporate dynamic, real-time simulation that restores the basic understanding of physical measurements~\cite{related_visceralization_2020} and transforms abstract data into an intuitive experience. This directly targets Level 2 SA (Comprehension).

\item \textbf{Support what-if analysis through computational steering.} To address the expert need for dynamic, hypothetical queries (Finding 3), the system must allow users to interactively modify simulation parameters and see the results immediately. This steering requires real-time computation and an intuitive input method whose affordances facilitate spatial understanding. The feedback for each modification should be visually apparent and easy to navigate. 
For example, incorporating third-person perspectives~\cite{related_sa_drone_2019} and overhead imagery~\cite{related_sa_whitlock_2020} can make the risks more concrete for Level 1 and 2 SA (Perception and Comprehension). 
Altogether, this mechanism enables immersive data exploration and directly supports Level 3 SA (Projection).

\item \textbf{Synthesize the analytical and intuitive modes.} The system's primary goal is to address the core challenge from Finding 3. It must not just contain these components, but synthesize them into a single, holistic environment. The analytical mode (symbolic data overlays) and the intuitive mode (embodied, visceral simulation) should be seamlessly integrated, allowing users to fluidly transit between analytical reasoning and embodied intuition for different analytical tasks.
\end{enumerate}

\section{LandSAR}

\begin{figure}[t]
    \centering
    \includegraphics[width=\linewidth, alt={The image shows the LandSAR system’s interaction between the physical and virtual worlds. The physical world comprises a base station, a VIVE Tracker for sensing, a Quest Pro MR display for recognizing gestures, and a terrain model. These physical components send data to the virtual world through a WebSocket server. The virtual world uses Unity to render landslide visualizations and simulations, which are then projected onto the terrain model, creating an immersive, tangible experience where the physical model enhances users' understanding of landslide risks.}]{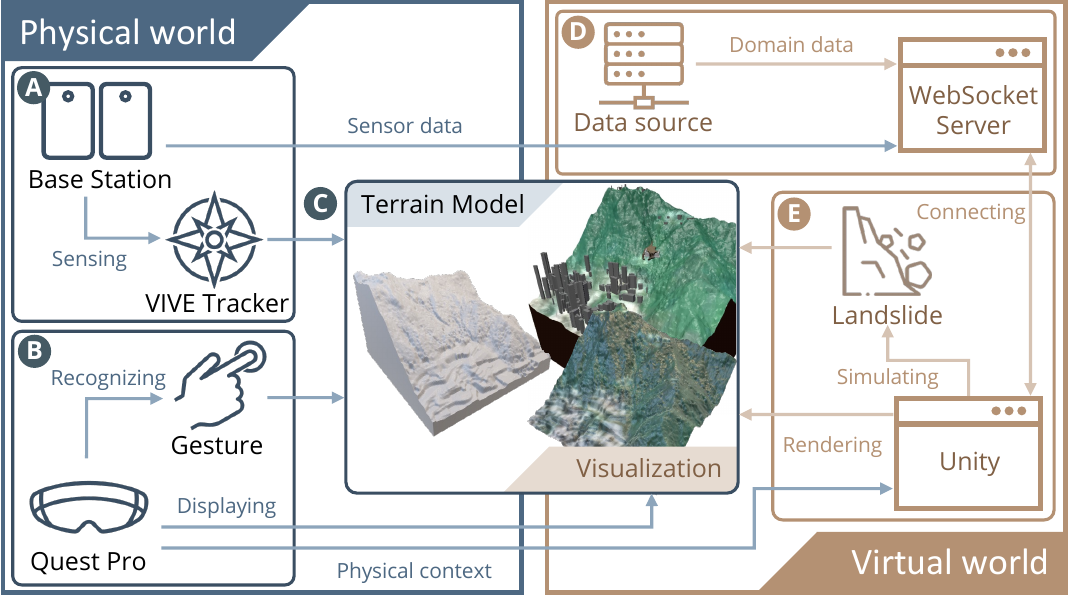}
    \caption{The system overview of LandSAR highlights the integration between the physical and virtual worlds. In the physical world, (A) the base stations and VIVE trackers collect sensor data from user interactions with a tangible terrain model, while (B) the Quest Pro MR display recognizes gestures and displays AR content. (C) The physical terrain model provides a foundation for these interactions and bridges the two worlds. In the virtual world, sensor data from the physical world is transmitted to a WebSocket server (D), enabling Unity to render visualizations and simulate landslides (E). These visualizations are aligned with the physical terrain model for enhanced user engagement and understanding.}
    \label{fig:system}
\end{figure}

\rw{
LandSAR is designed based on the requirements for tangible terrain models and situated analytics.
LandSAR consists of five main components that link the physical and virtual worlds (\cref{fig:system}).
These components work together to create an immersive and engaging experience for users (R3).
The tangible tracking system (A) supports reliable, real-time physical interaction with digital content (R2) without requiring users to learn a new set of mid-air hand gestures (R1).
The MR HMD (B) provides the capabilities of both AR and VR. Besides blending digital objects with physical contexts for in-situ interactions (R1), it allows users to experience the landslide immersively from a first-person perspective, increasing situated awareness (R4).
The terrain models (C) bridge the physical and virtual worlds, serving as the primary interfaces for geographic context and providing physical feedback to users.
Rendered visualizations are overlaid on and situated near the terrain models to support depth perception and haptic feedback during user interaction.
The WebSocket server (D) collects sensor data about users' movements from the tracking system and domain information from data sources to support analytical tasks and information retrieval.
The rendering system (E) simulates the dynamic environment to enhance users' situational awareness (R4). It renders visualizations according to physical interactions on the terrain models, ensuring that users can see the results of their actions in real time.
}

\subsection{\rw{Situated data visualization module}}
\begin{figure}[t]
    \centering
    \includegraphics[width=\linewidth, alt={The image shows LandSAR's situated visualization for landslide risk analysis. In Panel A, a user interacts with a terrain model while wearing an AR headset, exploring historical landslide events. Panels B to E are the view in the user's eyes. B displays a timeline of landslide occurrences, allowing users to select specific events. Panel C simulates potential landslide impacts under extreme rainfall conditions. Panels D and E show the ``causality view,'' where D presents rainfall records from the 2008 rainstorm and E shows the corresponding landslide susceptibility distribution. Together, these visualizations help users understand the historical and predictive aspects of landslide risks.}]{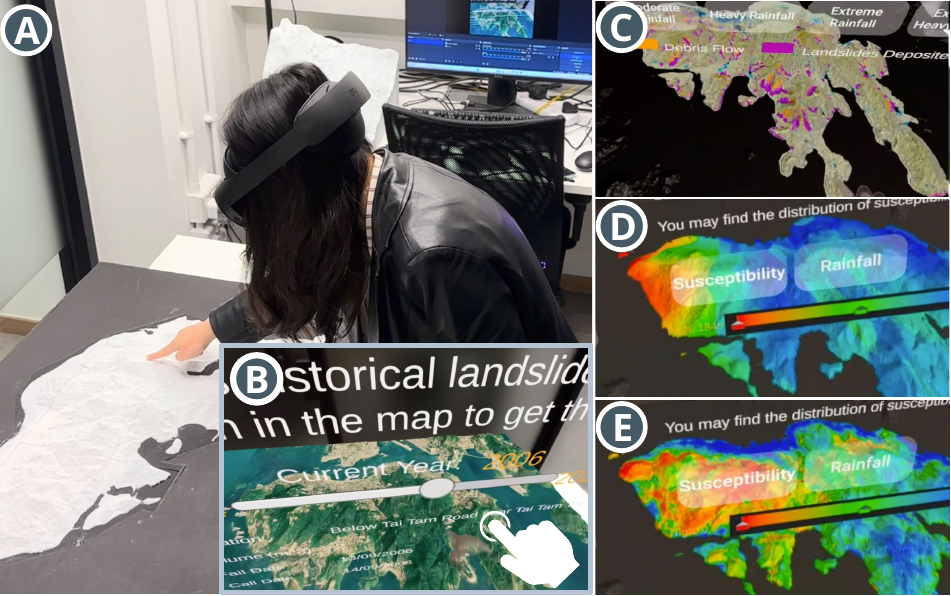}
    \caption{The situated visualization of LandSAR provides historical and predictive information about landslide risks. (A-B) History view shows the distribution of historical landslide events from 1984 to 2021. By touching the model, users can query specific events, such as the one shown from 2006. (C) Climate change view projects landslide impact under hypothetical extreme-rainfall scenarios~\cite {visual_prediction_zhou_2019}. (D-E) The causality view presents the historical rainfall records of the 2008 rainstorm event (D) and the corresponding landslide susceptibility distribution in the same event (E)~\cite{background_landslide_rainfall_2021}, providing a deeper understanding of the rainfall-landslide relationship.}
    \label{fig:map}
\end{figure}

Our situated visualizations are designed with three views on terrain models to guide users through three hierarchical phases of situational awareness: developing a comprehension of historical landslide patterns, building a foundational perception of landslide causes, and enabling prediction of future landslide risks under climate change.

The history view (\cref{fig:map}B) allows users to explore the historical landslide events from 1984 to 2021 on the map to build a deeper comprehension of their spatial and temporal patterns. 
To explore the data over time, users can perform temporal filtering using an interactive slider widget, with a specific focus on notable storms (\eg 2005 and 2008~\cite{background_landslide_rainfall_2021}).
The affected areas are highlighted to show the impact of landslides.
Each landslide is represented by a glyph positioned on the map. The glyph's visual properties, such as size, can be mapped to the scale of the landslide, allowing users to quickly identify significant events at a glance. Tapping on any glyph triggers an annotated panel displaying detailed information about that specific event, including its location, scale, etc.
This combination of an overview and on-demand detail supports a seamless exploration of past events.

The causality view (\cref{fig:map}D-E) immerses users in a historical rainstorm case study to establish a foundational understanding of landslide causes.
The rainstorm from 6-8 June 2008 was centered over Lantau Island and triggered more than 1,500 landslides on the island.
The view consists of two heatmap visualizations to facilitate direct comparison.
The first uses a heatmap to encode rainfall intensity, and the second shows landslide susceptibility during the same period.
To facilitate intuitive comparison, both visualizations employ a shared color mapping: a blue-to-red color scale encoding low values in blue and high values in red.
The strong visual correlation between the intense colors on the heatmap and the high-risk areas on the susceptibility map allows users to instantly perceive the fundamental relationship between heavy rainfall and landslide risk.

The climate change view (\cref{fig:map}C) visualizes the predicted impact of landslides in Hong Kong Island from simulation models.
This view facilitates what-if analysis by visualizing the predicted impact of future extreme rainfall events, using models by Zhou\etal~\cite{visual_prediction_zhou_2019} that simulate rainfall 1.5, 2.5, and 3 times stronger than the 2008 rainstorm.
The predicted impacts are rendered directly onto the terrain model. Debris flow paths are encoded using a sequential orange-to-red color scale to represent the flow's destructive potential, while areas of potential debris deposits are highlighted in purple for high contrast against the natural terrain. 

\subsection{\rw{Landslide simulation module}}
\begin{figure}[t]
    \centering
    \includegraphics[width=\linewidth, alt={The image shows a landslide simulation designed to improve situational awareness. Panel A provides a 3D model with the location of residential areas and a protective rigid barrier, while Panel D shows an analysis of the landslide’s runout. Panels B, C, E, and F display a real-time simulation of the landslide. Panel E demonstrates dynamic interaction by changing the barrier's location to alter the landslide’s course. Panel F offers a first-person perspective, showing the flow of debris towards the protected areas.}]{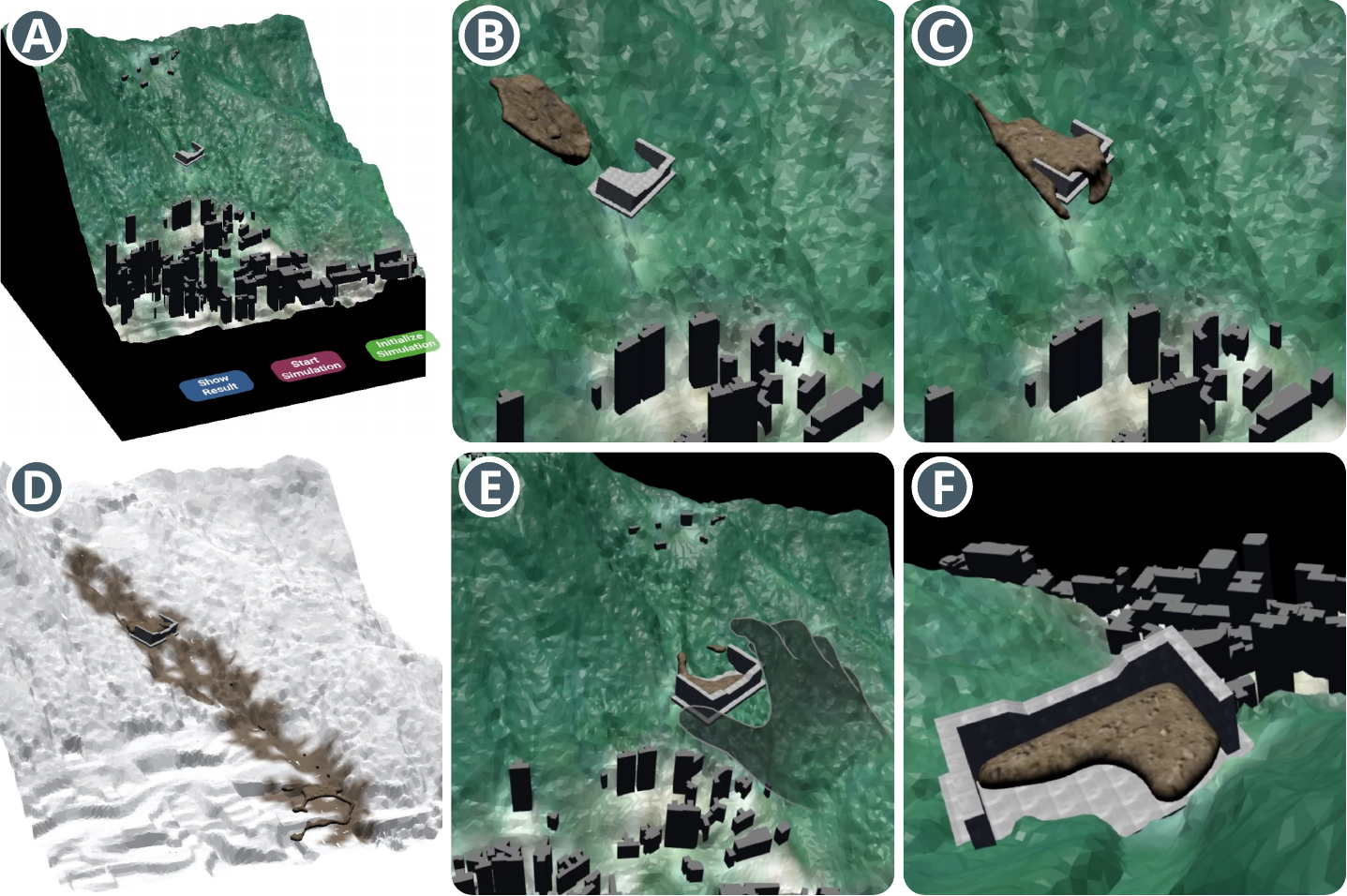}
    \caption{The landslide simulation is designed to enhance SA: \textit{Perception} with (A) understanding the residential areas and placing the rigid barrier for protection; \textit{Comprehension} with (B-C) real-time landslide simulation and (D) runout analysis; \textit{Prediction} with (E) dynamically changing barrier's location and (F) observing debris flows from a first-person perspective.}
    \label{fig:simulation}
\end{figure}
The landslide simulation is a core component of LandSAR. The simulation is situated in the PSR region. It transforms users from passive observers into active participants by leveraging tangible interaction for decision-making and real-time visceralization for building a deep, intuitive understanding of landslide dynamics.

\subsubsection{Preparation}
Before the landslide starts, users are encouraged to familiarize themselves with landslide precautionary signals (weather forecasts and landslide warnings).
As users might not be familiar with the region, they can set the scene to further understand the area's elements (\cref{fig:simulation}A).
Users can tap on the terrain models to toggle the visibility of virtual objects (\eg buildings and vegetative covers), or reference a landslide susceptibility visualization, helping them perceive high-risk zones and the infrastructure they need to protect.
We employ tangible interaction as the primary modality for this phase. 
Trackers can be used to move and orient rigid barriers on the terrain model, while their parameters (\eg height, width, angle) can be adjusted in the interface.
This direct manipulation grounds the decision-making process in a tactile experience, offloading the cognitive challenges of abstract spatial interaction.

\subsubsection{Simulation}
Debris flows are typically modeled with various deformation continuum mechanics, which require tremendous computational power.
To provide real-time simulation, we simplified the model with fluid- and particle-based simulation techniques. 
For realism, we consulted geotechnical engineers to understand the physical properties of landslides and ensure boulders are pushed by the fluid (\cref{fig:simulation}B).
Boulders can be easily caught by barriers, while fluid may still overflow if the volume surpasses the barriers' capacities or is poorly positioned (\cref{fig:simulation}C).
The benefit of using real-time simulation is that it allows users to correct design mistakes, such as a misplaced barrier.
The god hand~\cite{related_tangible_miniature_2021} in \Cref{fig:simulation}E controls the barrier's location with the tracker, and the simulated debris flows will be affected accordingly.

Once the barrier is placed, the simulation phase immerses the user to build comprehension through a visceralization of the landslide event. From a default third-person AR perspective, users observe the overall dynamics as debris flows and boulders interact with the terrain and their barrier (\cref{fig:simulation}B-C).
To enhance situational awareness and support visceralization, users can touch any point on the terrain model to instantly teleport to that location and observe the landslide from a first-person perspective (\cref{fig:simulation}F).
This provides an immersive sense of the landslide's velocity, allowing users to spot danger signs, such as a creaking barrier.
Following the design recommendations of Yang\etal~\cite{visual_design_yang_2021}, a bird's-eye view is attached to the left hand as a WIM widget (\cref{fig:teaser}C) to compensate for the loss of situational awareness in the first-person perspective (\ie in the field~\cite{related_sa_whitlock_2020}). Tapping the bird's-eye view returns users to AR mode.
The simulation runs in real time, allowing users to use the ``god hand'' to dynamically adjust their barrier mid-event, immediately see the consequences, and test their evolving understanding of flow mechanics.

\subsubsection{Analysis}
After the simulation, the final phase provides users with visual analytics tools to analyze the outcome and devise better strategies. The analysis begins with a comparative summary visualization showing the final debris flow trajectory with and without the installed barriers (\cref{fig:simulation}D). The comparison allows users to immediately perceive the overall effectiveness of their intervention.

To facilitate a deeper diagnostic, the system also provides a risk analysis based on the fundamental principle that:
\begin{equation}
  \label{eq:risk}
  Risk = Hazard \times Vulnerability
\end{equation}
where hazard is the physical impact of debris flows (\eg impact force $F$ and velocity $v$) and vulnerability represents the potential for loss in the affected area (\eg buildings and residents). Instead of directly showing users complex equations, LandSAR translates these components into two distinct, interactive visual layers on the terrain model.

In existing international design guidelines~\cite{visual_impactforce_2012,visual_impactforce_2014}, the impact force exerted by a debris flow is calculated as follows:
\begin{equation}
  \label{eq:impactforce}
  F = \alpha\rho v^2h_0w 
\end{equation}
where $\alpha$ is a dynamic impact coefficient ($\alpha$ is set as 2.5 for the rigid reinforced concrete barrier suggested by Kwan\cite{visual_impactforce_2012}); $\rho$ is the flow density; $v$ is the velocity of the flow; $h_0$ is the flow depth and $w$ is the initial width of the debris flow. In this formula, the impact force is proportional to the square of the velocity (v²). This means a small increase in speed can dramatically increase its destructive power.

With a barrier in place, the energy of the overflow is attenuated when the debris flow climbs over the barrier and lands back onto the terrain~\cite{visual_velocity_2021}. The new velocity can be calculated as follows:
\begin{equation}
  \label{eq:velocity}
  v_i = C_r(\theta)v_r
\end{equation}
where $v_r$ is the initial velocity of the debris flow and $v_i$ is the flow velocity toward residential areas. 
A landing coefficient $C_r = R\cos\theta$ is introduced, where $R$ is the reduction factor of landing velocity due to friction between the flow and the terrain and $\theta$ is the angle of the terrain slope.

Our system calculates this hazard potential along the landslide's trajectory and encodes it as a color-graded hazard map. Areas subjected to high impact forces are rendered in deep red, while areas with attenuated forces, such as those downstream of a barrier, are shown in yellow or green. This allows users to visually trace the energy of the landslide and see precisely where the barrier was most effective at reducing the physical threat.
To visualize vulnerability, we created a simplified formula that considers both building density and population in the area:
\begin{equation} 
\label{eq:vulnerability} 
V = w_b D_b + w_p D_p
\end{equation}
where $D_b$ is the building density, $D_p$ is the population density in the affected area, and $w_b$ and $w_p$ are their respective weighting factors.
By overlaying these two visual layers, users gain a comprehensive understanding of the overall risk. They can instantly identify high-risk areas where a high-energy hazard intersects with a highly vulnerable population. To investigate these areas further, users can perform tangible and spatial queries. Similar to Uplift~\cite{related_tangible_uplift_2021}, users can pick up barriers to inspect the information recorded in the simulation, such as flow rate and impact force.
Pointing to any location reveals specific statistics about the simulation, such as flow depth across timespans.
This multi-layered analysis, moving from a high-level visual comparison to a detailed, data-driven diagnostic, empowers users to not only see what has happened but to understand why, helping them predict potential consequences and devise more effective mitigation strategies.

\subsection{System implementation}
The LandSAR system is implemented in two core modules.
For the virtual world, we developed the project in Unity 2021.3.16f1.
The terrain models\footnote{\href{https://www.landsd.gov.hk/en/spatial-data/open-data/kf_dtm.html}{https://www.landsd.gov.hk/en/spatial-data/open-data/kf\_dtm.html}} and most of the domain data\footnote{\href{https://data.gov.hk/en-datasets/provider/hk-cedd}{https://data.gov.hk/en-datasets/provider/hk-cedd}} were obtained from the government's open-data initiative.
The 3D models of the buildings and regions were processed with Blender.
The fluid dynamics are modeled using a moving least squares material point method provided by Zibra Liquids\footnote{\href{https://zibra.ai/}{https://zibra.ai/}}.
This solution provides real-time landslide simulation that considers the terrain surface, building models, and designed intervention (\ie rigid barrier) as collision constraints.
To balance realism and performance, we simplified the geometric complexity of surrounding structures while preserving terrain features that are critical for collision handling and rendering.

For the physical setup, we employed the Meta Quest Pro as the AR display because of its enhanced field of view (FOV) and gesture recognition.
Since the real-time simulation requires more computational power than standalone MR HMDs can provide, we connected the Quest Pros to a desktop computer via a 6-meter cable supporting 5 Gbps bandwidth.
The desktop is equipped with an Intel i7 CPU clocked at 3.6 GHz, 32 GB RAM, and an RTX 3090 Ti GPU.
The tracking system is based on SteamVR base stations, set diagonally across the room, and HTC VIVE trackers.
It provides room-scale tracking during interactive sessions, allowing users to manipulate barriers with low latency while keeping interaction centered on the terrain model.
Due to technical privacy constraints, the implementation of the Quest Pro's passthrough mode conflicts with the SteamVR base stations.
Therefore, we calibrated the trackers' positions using a different device and sent their position and rotation information through a WebSocket server.
This information is then mapped to the corresponding objects in Unity to provide AR correspondence.
The trackers are used in two ways: (1) to locate the terrain model by placing them at predefined locations, and (2) to bind them to the virtual rigid barrier to support real-time landslide simulation.
This eliminates the need to use the official controllers, which can induce an additional learning curve for users without previous AR/VR experience.

\subsection{Tangible terrain models}
\begin{figure}[t]
    \centering
    \includegraphics[width=\linewidth, alt={The image displays two tangible terrain models. On the left side, a detailed scale model depicts a mountainous area with visible topography and urban zones. On the right side, a larger section represents Hong Kong Island. Both models are installed on an acrylic board.}]{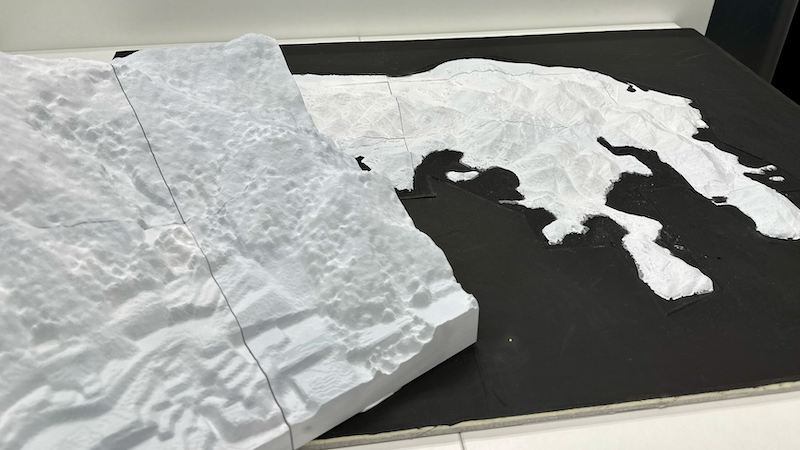}
    \caption{The tangible terrain model in LandSAR. Left: A scale model containing the mountainous and urban areas. Right: A scale model of Hong Kong Island, installed on an acrylic board.}
    \label{fig:object}
\end{figure}

\Cref{fig:object} shows the printed Hong Kong Island model, which we chose to provide an overview of this mountainous region, and the Po Shan Road (PSR) area, which refers to the serious landslide incident in 1972\footnote{\href{https://www.cedd.gov.hk/filemanager/eng/content_414/er229links.pdf}{https://www.cedd.gov.hk/filemanager/eng/content\_414/er229links.pdf}}.
We normalized the terrain models onto square grids and reduced the granularity by resampling them to remain compatible with the 3D printer's requirements.
The height of the PSR model is scaled by a factor of 1.5, as a domain rule of thumb suggested by domain experts to compensate for the perceptual insensitivity to slope elevation, which is also commonly adopted in scientific visualization~\cite{related_tangible_herman_2025}.
They were fabricated using an HP Jet Fusion 540 device with HP 3D High Reusability PA 12, a nylon-based material with high durability and good retention of fine detail.
The interior of the elevated terrain model is supported by pillars to save material, following a guideline inspired by Allahverdi\etal~\cite{object_landscaper_2018}.
Color-based perception could be weakened in immersive settings, mainly because of interference from the real-world background~\cite{visual_guide_whitlock_2020}.
Thus, we selected white material to prevent potential obfuscation of visual encodings.
Due to the size limit, the Hong Kong Island model was printed in 11 parts and installed on a $100 \times 75$ cm acrylic sheet.
The sheet's color is painted in black to highlight the Island model.
The PSR model was printed in four parts and not bonded, allowing future interaction designs.

\section{Evaluation}

\begin{figure}[t]
    \centering
    \includegraphics[width=\linewidth, alt={The image shows a workshop setting where participants engage with the LandSAR system. A user interacts with a tangible interface to modify the position of protective barriers on a terrain model. The user is experiencing a first-person view of a landslide simulation using an AR headset. Another image shows that users are seen interacting with the system, both with and without the use of the tangible terrain model, showcasing different ways of engaging with the visualizations and simulations during the session.}]{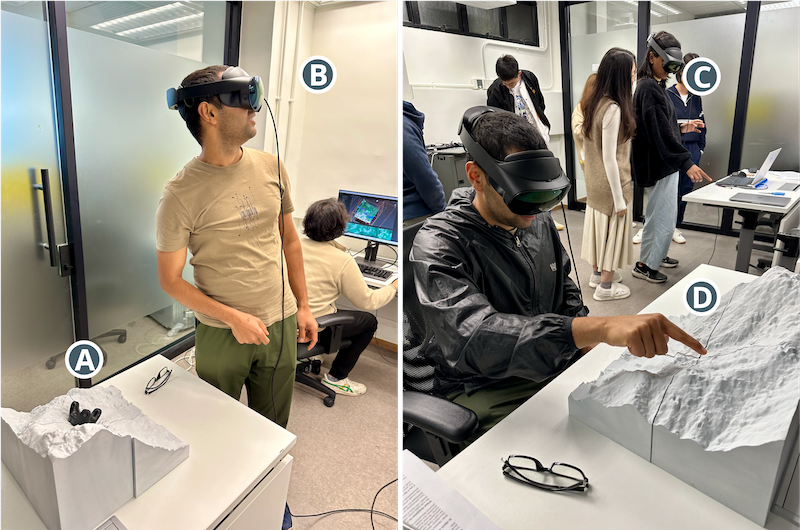}
    \caption{The setting of workshop sessions. 
    (A) Modifying the position of barriers with tangible interfaces. 
    (B) Experiencing the first-person perspective during a landslide simulation. 
    (C) and (D) Interacting with and without a tangible terrain model.
    }
    \label{fig:case_one}
\end{figure}

Following the co-design methodologies, we organized monthly feedback sessions with domain experts to assess the system designs and gather constructive feedback.
To evaluate the final prototype, we conducted a two-part study: (1) a formative workshop with experienced researchers and graduate students ($N=12$) to validate the system's design and usability, and (2) a summative expert interview with senior geotechnical engineers ($N=3$) to evaluate its impact on expert-level Situational Awareness (SA) and its effectiveness in synthesizing the analytical and intuitive modes.

\begin{figure}[t]
    \centering
    \includegraphics[width=\linewidth, alt={Baseline Landslide Situation Awareness Questionnaire results. The chart shows the 5-point Likert scale (1=negative to 5=positive) responses of the 12 participants, assessing their self-reported knowledge in five categories: Perception, Comprehension, Prediction, Preparedness, and Confidence.}]{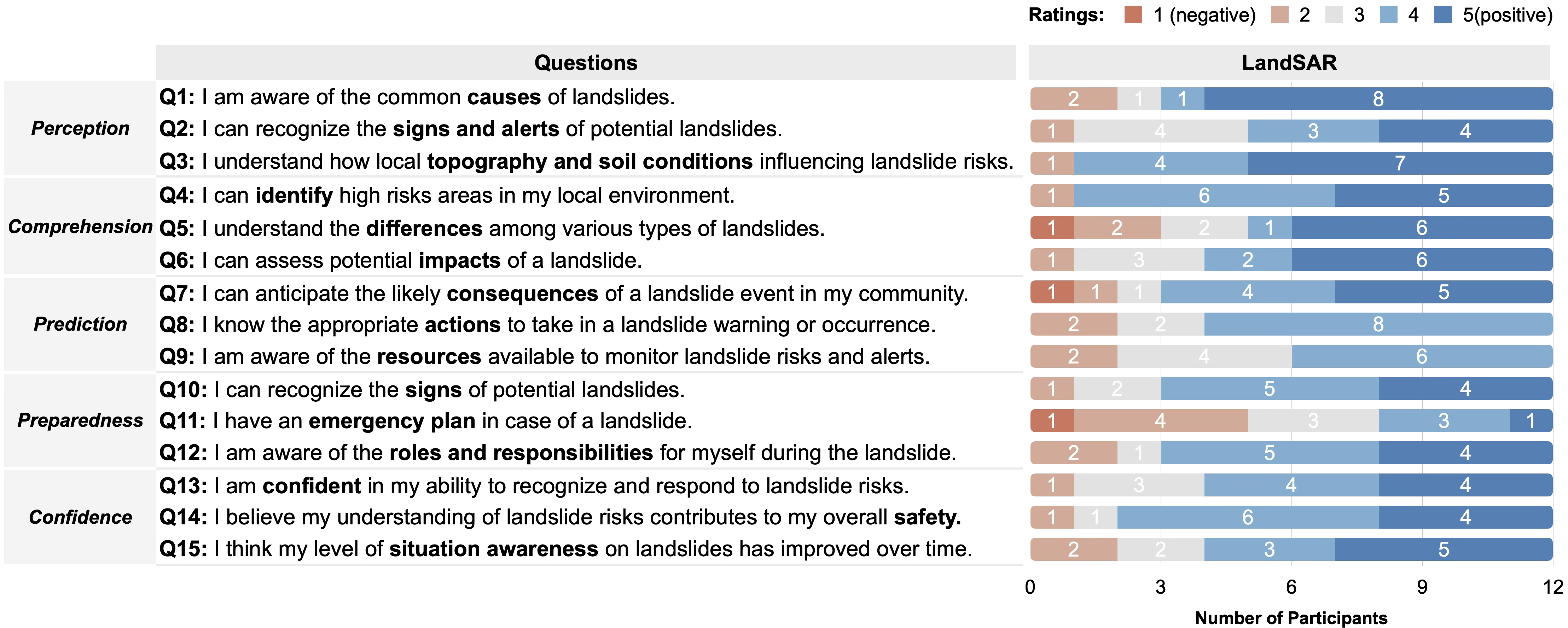}
    \caption{Self-reported baseline situational awareness for landslide.} 
    \label{fig:baseline}
\end{figure}

\subsection{Setup and participants}
We recruited 12 participants (P1--P12) with a variety of academic backgrounds from the local university through social media, consisting of graduate students and junior researchers in civil engineering (4 females, 8 males; aged 24--35).
They were not compensated.
Their baseline situational awareness regarding landslide scenarios (\cref{fig:baseline}) was self-reported using a 5-point Likert-scale questionnaire across five dimensions: perception (Mean = 4.17), comprehension (4.03), prediction (3.58), preparedness (3.53), and confidence (3.97).
The participants were divided into three groups and took part in workshop sessions lasting 90-120 minutes.
The workshops were facilitated by five interdisciplinary co-authors, who provided technical support and answered domain-specific questions.

To enhance the efficiency of the workshops, we separated the situated visualization and landslide simulation modules, running both concurrently on two Quest Pros placed side-by-side. 
Despite being designed for standalone use, the situated visualization was connected to a laptop for added computational support and streaming capabilities. 
The laptop is equipped with an Intel i9 processor clocked at 2.5 GHz, 32 GB RAM, and an RTX 3070 Ti GPU.
This setup ensured smooth operation, reducing the need for repeated explanations and minimizing waiting time for participants.

\begin{figure}[t]
    \centering
    \includegraphics[width=\linewidth, alt={The image presents the User Experience Questionnaire (UEQ) results for the LandSAR system. Each row contrasts opposing adjectives, such as Obstructive vs. Supportive and Complicated vs. Easy. The color gradient represents ratings from 1 (red) to 7 (dark blue), with numbers inside the bars indicating how many users gave each score. The system generally received higher marks, with many users rating it as supportive, efficient, clear, exciting, interesting, inventive, and leading edge, though there were some lower ratings for complexity.}]{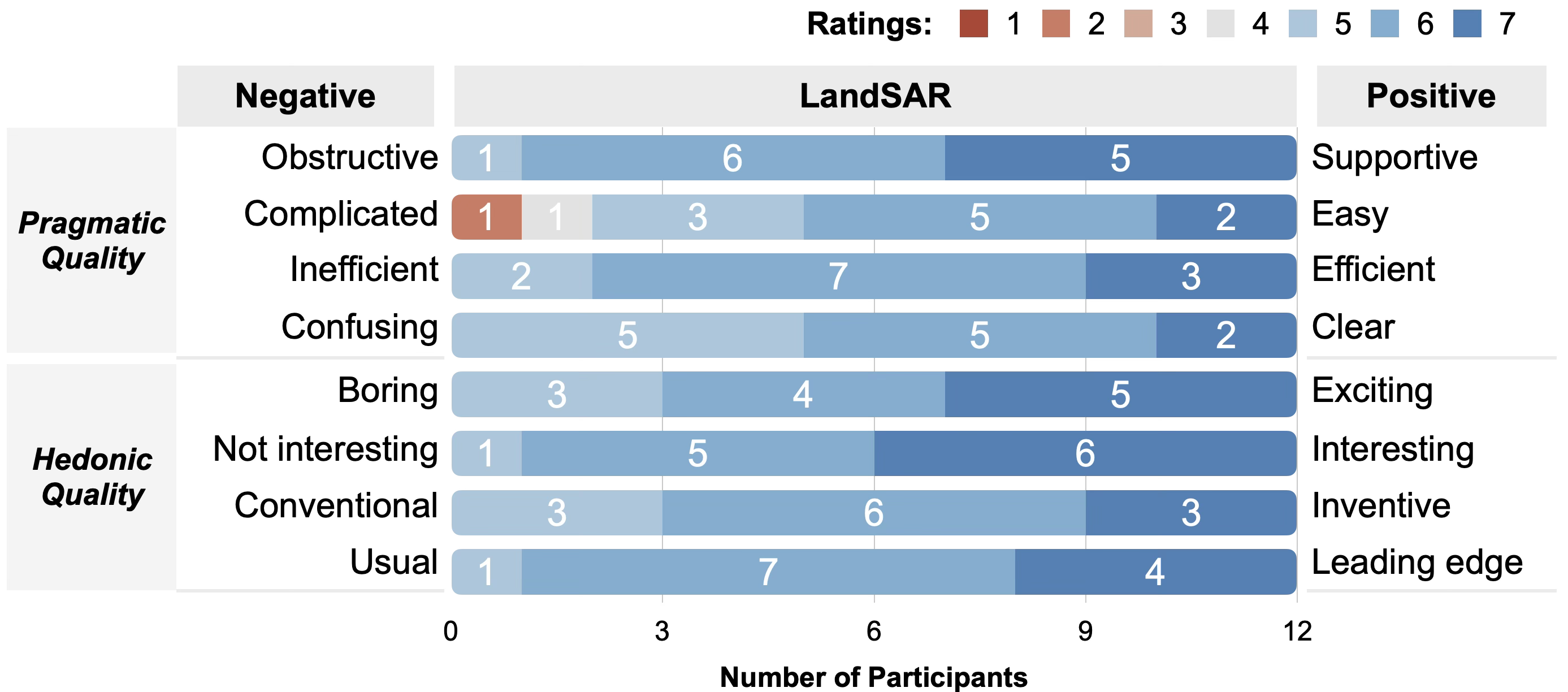}
    \caption{The User Experience Questionnaire (UEQ) results in rating LandSAR by the overall system usability and user experience. 
    }
    \label{fig:ueq}
\end{figure}

\begin{figure}[t]
    \centering
    \includegraphics[width=\linewidth, alt={The image presents the Situation Awareness Rating Technique (SART) results for LandSAR, measuring users' perceived situational awareness. The coding is the same as in Figure 8. The results show a range of perceptions, with some users feeling that the system was stable and simple, while others found it more complex and alert, indicating diverse levels of situational awareness.}]{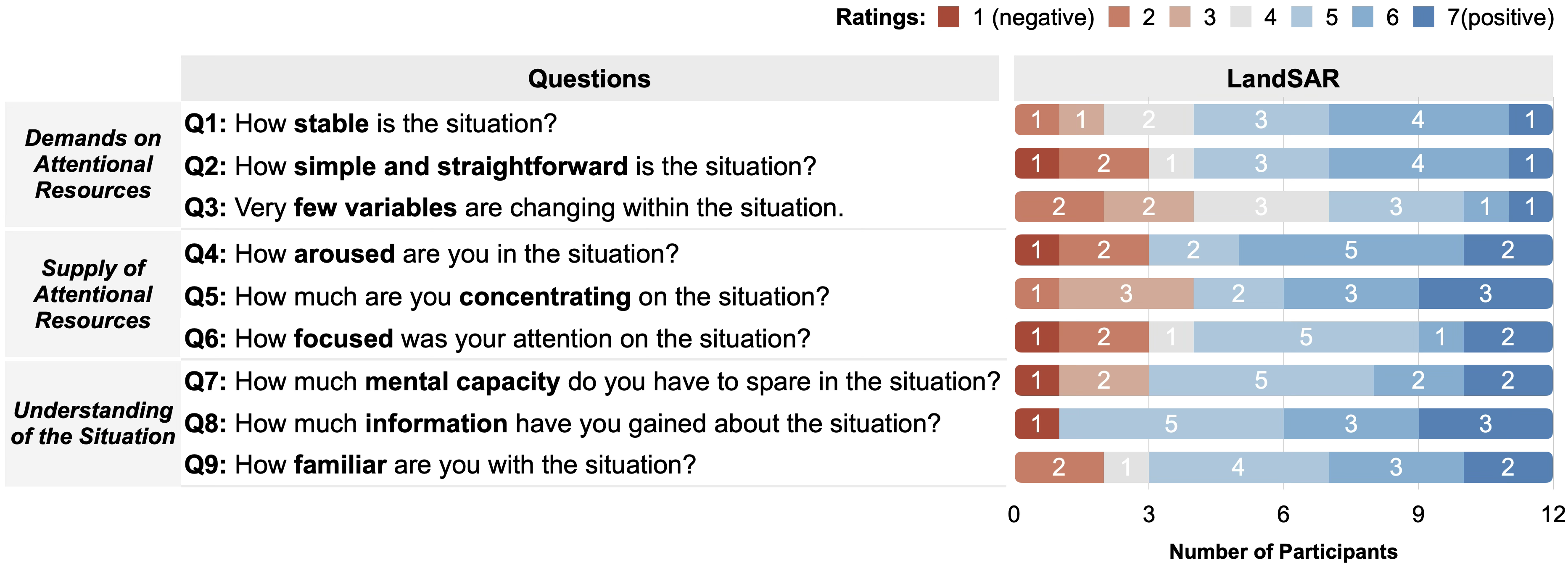}
    \caption{Situation Awareness Rating Technique (SART)~\cite{taylor2017situational} results for the situational awareness induced by LandSAR.
    }
    \label{fig:sart}
\end{figure}

At the beginning of each session, participants were invited to fill in consent forms, a demographic questionnaire, and the landslide situational-awareness questionnaire to assess their background.
After that, the session hosts introduced the system objectives and functions, demonstrating the individual components.
The Oculus internal casting system was used for situated visualization, while Unity's game view was used for real-time simulation. Since AR passthrough was not available in Unity play mode because of privacy restrictions, the simulation was displayed through the demonstrators' view.
Participants were then guided through the system components and experienced the prototype hands-on.
Participants were encouraged to explore the same components with and without the tangible terrain model to experience the differences.
Once all participants completed their hands-on experience, they were invited to fill in two questionnaires: the User Experience Questionnaire (\cref{fig:ueq}) and the Situation Awareness Rating Technique~\cite{taylor2017situational} (\cref{fig:sart}).
These questionnaires were used to evaluate their interaction with LandSAR.
Lastly, a semi-structured group interview was conducted, in which participants shared feedback on the system's design.
The interviews were audio-recorded and transcribed for analysis.
The qualitative data was analyzed using thematic analysis to identify the key insights regarding situational awareness and the synthesis of analytical and intuitive modes reported in the findings.
For the expert interview, three senior geotechnical engineers (E1--E3) from the same department referenced in \cref{sec:interview} were invited to a two-hour interview.
Given the limited time available for both setup and the interview, only the PSR model was demonstrated at the government building. 
The ``situated visualization'' was demonstrated without the tangible terrain model.
The rest followed the design used in the workshops, except only the laptop was present for computational support.

\subsection{Feedback on LandSAR}
Participants provided valuable feedback and insights for the ongoing development of LandSAR. \rw{Below, we summarize key themes and extend them with additional observations and expert comments from the workshops and follow‑up interviews.}

\textbf{First-person perspective offers the greatest impact.}
\rc{The first-person perspective was consistently described as the most memorable and impactful feature of the system. 
The transition from a bird's-eye AR overview around the terrain model into an egocentric, on-the-slope view produced a strong immediate impression and changed participants' perceived role from observers to stakeholders. 
P5 remarked, ``Nobody wants a barrier the size of a swimming pool blocking natural views... LandSAR lets me preview how these designs would appear from a resident's point of view.'' 
P2 echoed, ``The first-person perspective changes my role from a detached observer to an engaged stakeholder.''
Experts also emphasized that this perspective is particularly valuable because real-world site reconnaissance is mandatory but logistically constrained.
Some vantage points are unsafe or inaccessible, especially after a major landslide event. 
They saw value in using LandSAR to ``virtually stand'' at locations that would be too dangerous or impossible to reach in the field, and to rehearse how a barrier or landslide would be experienced.}

\rc{At the same time, the first-person view exposed several perceptual and technical challenges. First, the coarse terrain mesh used for real-time simulation became very salient when viewed up close. Experts reported that standing on the slope, they could clearly see large triangular facets but could not intuitively judge their physical size. One facet might represent 5-10 m in reality, which corresponds to several levels of a building. Without that domain knowledge, participants struggled to understand the true scale and steepness of the slope from the first-person view. They suggested two complementary remedies: 1) \textit{Decoupling simulation and rendering meshes.} A coarse mesh could still be used for real-time simulation, while a finer terrain mesh (or high‑resolution texture) is used purely for visualization. In their view, this would preserve performance while restoring a more believable perception of surface detail close to the user. 2) \textit{Adding size references.} Several participants explicitly asked for reference objects such as human figures, trees, and buildings along the flow path to ``anchor'' the perceived size of the landslide. Experts also recommended visualizing damaged or collapsed buildings when debris hits them, arguing that visible consequences would reinforce users' sense of danger and better convey the severity of different design choices.}
\rc{Second, some participants reported mild dizziness when rotating quickly in first-person view, especially when passthrough video and PC streaming were active simultaneously. Experts noted that in this mode the physical room is almost irrelevant as they are focused on the virtual slope and debris. They would prefer a fully virtual (opaque) background if that would yield smoother motion. In contrast, they preferred passthrough in the third-person exocentric view, where seeing the physical terrain model and surroundings reinforced the feeling of observing the event ``from outside.'' This suggests a design guideline: \textit{use full VR for first-person visceralization, and mixed reality for third-person overview.}}

\textbf{The workflow synthesizes analytical and intuitive understanding.}
All participants (P1-12) and experts (E1-E3) appreciated the simulation workflow for its comprehensive approach. They agreed that the simulation effectively covered crucial aspects such as identifying warning signs (Level 1 SA), understanding landslide mechanics (Level 2 SA), and predicting potential runout areas (Level 3 SA). 
For expert analysis, participants and experts confirmed the situated visualizations allowed them to ``engage directly with the terrain model and link various data visualizations, such as rainfall patterns and soil characteristics, to the landscape.'' The experts noted this synthesis was a primary strength. E2 stated, ``The ability to see the simulation on the physical model while also having the analytical data is what we've been missing. I can test my hypothesis and immediately see the simulated consequence, which builds trust in the model.'' This feedback is consistent with our research goal of synthesizing the analytical and intuitive modes.
They proposed that historical events could be simulated side-by-side, with and without barriers, illustrating the effectiveness of protective actions and helping to raise awareness about the mitigation strategies.

\textbf{Bridging the gap in landslide preparedness.}
In the baseline questionnaire results (\cref{fig:baseline}), a notable discrepancy exists between high-level conceptual understanding and specific perception. While participants reported high awareness of the ``common causes'' of landslides (Q1) and general ``topography and soil conditions'' (Q3), their confidence was more varied when asked to ``identify high risks in my local environment'' (Q4) or recognize specific ``signs and alerts'' (Q2). Furthermore, we identified a clear theory--practice gap: a strong majority (8 out of 12) reported knowing the ``appropriate actions'' to take (Q8), but 5 participants (42\%) rated themselves negatively (1 or 2) on having an ``emergency plan'' (Q11).
These baseline findings suggest that participants' high overall confidence (Q13, Q14) may stem from a false sense of security, disconnected from actionable preparation. LandSAR is designed to directly address this by transforming abstract risk concepts into a tangible, localized, and personal experience, thereby bridging the gap between passive knowledge and active situational awareness.

\textbf{Ethical considerations for stakeholder communication.} The experts (E1--E3) raised important ethical considerations. When terrain models closely mirror actual geographic locations, they can be ``easily recognized by stakeholders.'' Experts cautioned that this high fidelity, while essential for their analysis, creates a communication challenge. Pinpointing specific locations could lead to a false sense of security (if an area is shown as ``safe'') or cause undue alarm. E1 explained, ``When stakeholders see familiar places, they will ask if they are really at risk. This tool increases our responsibility for data communication.'' This finding does not diminish the tool's value for expert analysis. Instead, it highlights a critical design consideration for a new class of tools: how to effectively transition from a high-fidelity expert analysis tool to a responsible stakeholder communication tool.

\section{Discussion}

Our findings demonstrate that LandSAR's value lies in effectively synthesizing the analytical and intuitive modes. This synthesis creates a holistic environment where each component maps to a different, complementary level of Situational Awareness (SA).

\subsection{Lessons learned}

\textbf{Mapping design components to situational awareness.} Our core research question explored how to synthesize analytical reasoning and embodied intuition. We found the components of LandSAR map directly to the three levels of SA:

\begin{itemize}[noitemsep,topsep=0pt,label=$\diamond$]
\item Analytical overlays for perception: Situated AR visualizations facilitate the analytical mode for abstract data and were most effective for establishing Level 1 SA (Perception). By projecting data like risk zones, rainfall intensity, and soil characteristics onto the model, experts could immediately perceive the current state of the terrain and answer ``what'' and ``where'' questions.
\item Simulation-based visceralization for comprehension: The real-time simulation facilitates the intuitive mode and improves Level 2 SA (Comprehension). Instead of just perceiving a static risk map, experts could see the physical mechanics of the flow. This simulation acts as an experiential medium, transforming abstract fluid impact analysis into an intuitive understanding of ``how'' and ``why'' the landslide behaves as it does.
\item Computational steering for projection: The interactive ``what-if'' capability of computational steering was the primary driver for Level 3 SA (Projection). By tangibly placing a barrier and immediately seeing its consequences, experts could test hypotheses and project the future status of the terrain. When errors occurred, they could correct them through precise tangible interactions.
\end{itemize}

\textbf{Data physicalization as a cognitive and perceptual anchor.} A key finding from our formative study was the ``disembodied gap'' of mid-air gestures, which caused cognitive load and perceptual ambiguity. The high-fidelity 3D-printed terrain model solved this in two ways. 
As a perceptual anchor, it provided a stable, passive-haptic referent that grounded the user's perception and reduced \rc{spatial ambiguity}.
As a cognitive anchor, the high-fidelity model provided an unambiguous ground truth for the terrain's shape. As E2 noted, when looking at satellite images, ``sometimes your eyes deceive you with the varying shadows and lighting.'' The physical model allowed experts to confirm and clarify complex features like troughs, ridges, and valleys that might be misinterpreted visually.

\textbf{Simulation as a low-cost verification tool.} The most useful ability enabled by this synthesis is its role as a what-if tool for design verification. In the real world, testing a mitigation strategy, such as the placement of a rigid barrier, is an expensive and irreversible engineering decision. LandSAR provides a ``cost-free'' environment to test these decisions repeatedly. This transforms the expert's workflow from one of static calculation to dynamic exploration. Experts could place a barrier, run the simulation, and look for its failure state. This immediate, visceral feedback allows them to build an iterative, vivid understanding of the design's flaws and fix them. This is physically and financially impossible in the real world. 

\textbf{The role of perspective on SA.} The ability to switch perspectives was critical, as each view served a different SA function:

\begin{itemize}[noitemsep,topsep=0pt,label=$\diamond$]
\item Overview (exocentric) perspective: The default AR view was essential for the analytical tasks. It provided the overview necessary for SA Level 1 (Perception) and Level 2 (Comprehension) of spatial relationships.

\item First-person (egocentric) perspective: This immersive view was essential for the embodied tasks. It allowed experts to understand the consequences of their design from a human-scale, stakeholder perspective. As P5 noted, it helps evaluate ``aesthetic concerns'' and the ``resident's point of view.'' This view makes the data visceral and personal, bridging the gap between abstract analysis and real-world impact.
\end{itemize}

\subsection{Limitations and future work}
\textbf{Simulation fidelity.} The current simulation \rc{uses a simplified Moving Least Squares Material Point Method (Zibra Liquids) rather than high-fidelity geotechnical rheology.} While effective for this evaluation, the specific geotechnical properties of debris flows (\eg soil viscosity, particle-fluid mixtures) are intrinsically different from simple fluids. \rc{This trade-off is necessary to achieve real-time responsiveness. For Level 3 Situational Awareness (Projection), the ability to perform rapid ``what-if'' analyses (Computational Steering) in real-time is considered more critical than engineering certification-level fidelity, which would require offline computation.} Future work should integrate more sophisticated, domain-specific debris flow models. 

\textbf{Computational performance.} The computational demand of running a real-time simulation alongside a high-fidelity AR passthrough remains a challenge. While LandSAR has barely noticeable latency in most cases, we observed that rapid head movements could induce latency and, as some participants reported, ``dizziness.'' Future HMDs with improved performance may mitigate this. Furthermore, our interaction relied on headset-based hand-tracking. This was sometimes compromised by occlusion, and it is not ideal for collaborative scenarios. Future systems should explore external-tracking systems to more robustly track multiple users and tangible proxy objects.

\textbf{Gaps in temporal control and analysis.} Our evaluation also highlighted a key area for future work: temporal control. The current simulation is limited to ``play'' and ``pause.'' Future systems should provide full procedural temporal control, including timepoint traversal and speed controls. This would allow an expert to isolate the exact moment a barrier fails, or examine a critical event backward and forward to build a deeper comprehension (SA Level 2) of its cause. Furthermore, our system is limited in simulating a single event at a time. Future systems can consider supporting temporal pattern analysis, such as ``show all historical events that overtopped this barrier.'' 

\textbf{Static vs malleable physicalization.} A primary design choice for LandSAR was the use of a static, high-fidelity 3D-printed terrain model. This provides a level of detail and haptic accuracy that malleable-medium systems like sandboxes cannot. However, our physicalization is non-mutable, meaning that experts can simulate on the terrain, but cannot change the terrain itself (\eg to dig a mitigation channel). \rc{In landslide risk management, the terrain topography is relatively static, while risk variables (rainfall, barriers, flow paths) are dynamic. Thus, a single physical print serves as a reusable ``canvas'' for infinite simulation scenarios.} Nonetheless, this creates a clear and exciting opportunity for future work: exploring a hybrid system that combines static, high-fidelity models with dynamic, shape-changing displays (\eg PolySurface~\cite{discussion_polysurface_2017}) to create a true digital twin system for both simulation and terrain-modification scenarios.
\section{Conclusion}

In this study, we addressed the critical challenge of enhancing expert Situational Awareness (SA) for complex, spatio-temporal landslide analysis. We characterized the cognitive challenge that arises when experts must mentally reconstruct dynamic physical processes from abstract, symbolic data, a process that impedes Comprehension (Level 2 SA) and Projection (Level 3 SA). We explored how the rational, analytical mode of immersive visualization and the intuitive, embodied mode of physical interaction can be effectively synthesized to create a holistic analytical environment. LandSAR integrates three components to bridge this gap: (1) data physicalization in the form of 3D-printed terrain models that act as tangible, haptic anchors to ground perception; (2) AR-based situated visualizations that supply the cognitive, analytical data overlays; and (3) simulation-based visceralization, which provides real-time simulation as an intuitive, embodied experience.

\acknowledgments{%
The authors would like to thank the anonymous reviewers for their constructive and insightful comments. This work was partially supported by HK RGC GRF grant (16214623) and Areas of Excellence Scheme (AoE/E-603/18).
}

\bibliographystyle{src/abbrv-doi-hyperref-narrow}

\bibliography{reference}

\end{document}